\documentclass[runningheads]{llncs}
\usepackage[T1]{fontenc}
\usepackage{graphicx}
\usepackage{booktabs}
\usepackage{graphicx}
\usepackage{adjustbox}
\usepackage{multirow}
\usepackage{float}
\usepackage{amsmath}
\usepackage{amssymb}
\usepackage{hyperref}

\usepackage{xcolor}
\usepackage{makecell}
\usepackage{tabularx}
\usepackage[table]{xcolor}

\let\originaluparrow\uparrow
\let\originaldownarrow\downarrow

\renewcommand{\uparrow}{\textcolor{green!50!black}{\originaluparrow}}
\renewcommand{\downarrow}{\textcolor{red}{\originaldownarrow}}

\begin{document}
\title{
BRIQA: Balanced Reweighting in Image Quality Assessment of Pediatric Brain MRI}
\author{
Alya Almsouti\thanks{Equal contribution.} \and
Ainur Khamitova$^{*}$ \and\\
Darya Taratynova$^{*}$ \and
Mohammad Yaqub
}

\authorrunning{Almsouti et al.}

\institute{
Mohamed bin Zayed University of Artificial Intelligence (MBZUAI), Abu Dhabi, UAE \\}

\maketitle              % typeset the header of the contribution
\begin{abstract}
Assessing the severity of artifacts in pediatric brain  Magnetic Resonance Imaging (MRI) is critical for diagnostic accuracy, especially in low-field systems where the signal-to-noise ratio is reduced. Manual quality assessment is time-consuming and subjective, motivating the need for robust automated solutions. In this work, we propose BRIQA (Balanced Reweighting in Image Quality Assessment), which addresses class imbalance in artifact severity levels. BRIQA uses gradient-based loss reweighting to dynamically adjust per-class contributions and employs a rotating batching scheme to ensure consistent exposure to underrepresented classes. Through experiments, no single architecture performs best across all artifact types, emphasizing the importance of architectural diversity. The rotating batching configuration improves performance across metrics by promoting balanced learning when combined with cross-entropy loss. BRIQA improves average macro F1 score from 0.659 to 0.706, with notable gains in Noise (0.430), Zipper (0.098), Positioning (0.097), Contrast (0.217), Motion (0.022), and Banding (0.012) artifact severity classification.  The code is available at https://github.com/BioMedIA-MBZUAI/BRIQA.
\keywords{Low-field MRI \and Quality Assessment \and MRI Artifacts \and Gradient-Based Reweighting \and Class Imbalance}
\end{abstract}
%
%
%------------------------------------------------------------------------
\section{Introduction}
% paragraph 1:  MRI importance for children + disadvantages of high field mri
% paragraph 2: Low field Mri advantages + the need fo automatic QA (add images)
% paragraph 3: previous works + Bigger models dont always mean better
Brain Magnetic Resonance Imaging (MRI) is an essential imaging modality to study pediatric brain development. In the early postnatal period, the human brain undergoes rapid growth and structural development; therefore, capturing these changes is important for improving our understanding of brain maturation and allowing the early detection of neurodevelopmental conditions \cite{introduction_Mri_in_neurodevelopment_2_and_non_ionizing,challenge_file,introduction_Mri_in_diagnosis,introduction_Mri_in_neurodevelopment}. While MRI is considered safe due to the absence of ionizing radiation \cite{introduction_Mri_in_neurodevelopment_2_and_non_ionizing}, high-field systems produce loud noise and require children to remain still in enclosed spaces for extended periods, often needing sedation, which is not ideal. In addition, these systems' high cost and maintenance requirements limit their accessibility in low- and middle-income countries.

\begin{figure}[t]
  \centering

  \begin{minipage}[t]{0.32\textwidth}
    \centering
    \includegraphics[width=\linewidth]{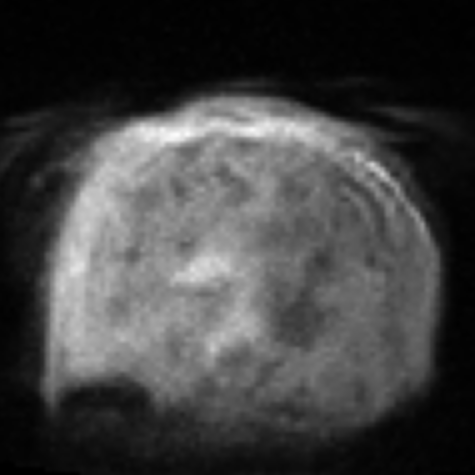}\\
    \textbf{(a)} Coronal view: Positioning, Motion, Contrast, and Distortion artifacts.
    \label{fig:coronal}
  \end{minipage}
  \hfill
  \begin{minipage}[t]{0.32\textwidth}
    \centering
    \includegraphics[width=\linewidth]{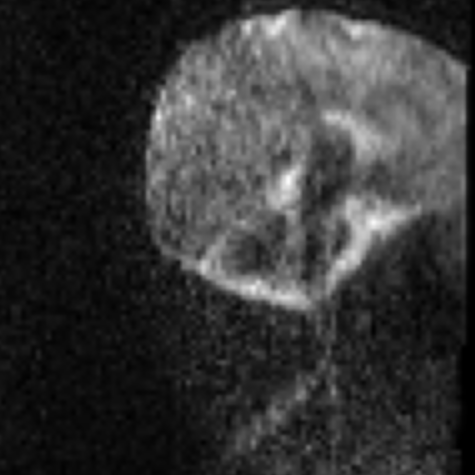}\\
    \textbf{(b)} Sagittal view: Noise, Zipper, and Positioning artifacts.
    \label{fig:sagittal}
  \end{minipage}
  \hfill
  \begin{minipage}[t]{0.32\textwidth}
    \centering
    \includegraphics[width=\linewidth]{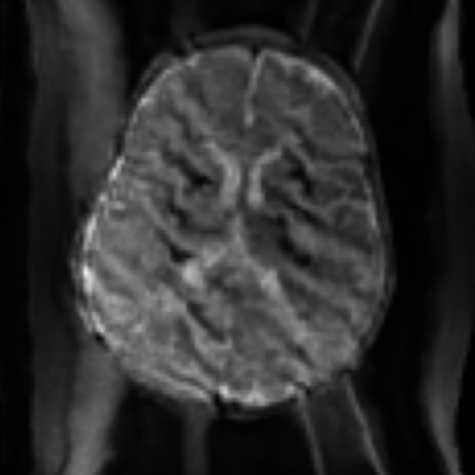}\\
    \textbf{(c)} Axial view:\\ Zipper and Banding artifacts.
    \label{fig:axial}
  \end{minipage}
  
  \caption{Scans from multiple patients obtained with the 0.064T Hyperfine SWOOP system, showing severe artifacts in different anatomical planes.}
  \label{fig:mri_scans}
\end{figure}

To address this, low-field MRIs offer an alternative solution with portable, point-of-care systems, reduced cost, quieter scans, and open designs, eliminating the need for sedation. However, the decreased signal-to-noise ratio in low field MRI poses limitations on the acquired image quality \cite{introduction_lmri_dis_and_advantages}, and introduces artifacts, as shown in Figure \ref{fig:mri_scans}. This makes image quality assessment essential to ensure that images meet specific standards and support diagnostic reliability, and given that manual Image Quality Assessment (IQA) is time-consuming and costly, automated solutions are crucial. This motivates Task 1 of the LISA Challenge 2025, the automatic assessment of the quality of the MRI scan in seven artifact classes.

Previous efforts have been made in brain MRI IQA, including the machine learning approach by Sanchez et al. \cite{FetMRQC}, which extracts image quality metrics from fetal brain MRI for automatic quality assessment. Deep learning approaches include Zhang et al. \cite{Zhang2024}, who proposed jointly segmenting the brain and assessing quality in fetal MRI slices, while Lou et al. \cite{lou2024noref} developed a contrastive learning method to enhance feature extraction, leveraging both spatial and frequency representations for quality assessment. The previous LISA 2024 Challenge \cite{challenge_file} featured Kim et al. \cite{kim2025axisguided}, who predicted scan orientation alongside quality assessment, Sundaresan et al. \cite{sundaresan2024automatedqualityassessmentusing}, who suggested synthesizing artifacts, and Zhu et al. \cite{zhu2025mambaout}, who developed a multi-label model combining gated CNNs and an ML-Decoder.

However, these prior studies rely on a single model architecture for all artifact types. In practice, performance can vary depending on the medical application; for example, ResNet may outperform DenseNet in some scenarios and vice versa \cite{Densenet_better_than_resnet,resnet_better_than_densenet}. Moreover, larger models do not necessarily perform better, particularly on small datasets \cite{bigger_not_better}. Therefore, it is beneficial to leverage diverse architectures of varying sizes, as different models may excel at identifying different artifact types based on their distinct visual patterns.

In this work, we introduce BRIQA, a method for the automatic assessment of artifact severity of MRI scans. BRIQA features a tailored model architecture for each artifact type, along with a gradient-weighting strategy and a custom batching technique to address class imbalance. The remainder of this paper is organized as follows: Section 2 describes the dataset used and  BRIQA framework including gradient-based reweighting and rotating batching, Section 3 and 4 presents experimental results with discussion, followed by conclusion at Section 5. 
% - Gradient-based class reweighting for severity classification:
% Applied a dynamic reweighting strategy that computes class-specific gradient norms and adjusts their contributions during training, effectively addressing severity-level imbalance.

% - Rotating fixed-ratio batching strategy:
% Proposed a novel rotating batching method that ensures class-balanced training batches with epoch-wise variation, promoting diversity and stability in learning under imbalanced settings

% \begin{table}
% \caption{Table captions should be placed above the
% tables.}\label{tab1}
% \begin{tabular}{|l|l|l|}
% \hline
% Heading level &  Example & Font size and style\\
% \hline
% Title (centered) &  {\Large\bfseries Lecture Notes} & 14 point, bold\\
% 1st-level heading &  {\large\bfseries 1 Introduction} & 12 point, bold\\
% 2nd-level heading & {\bfseries 2.1 Printing Area} & 10 point, bold\\
% 3rd-level heading & {\bfseries Run-in Heading in Bold.} Text follows & 10 point, bold\\
% 4th-level heading & {\itshape Lowest Level Heading.} Text follows & 10 point, italic\\
% \hline
% \end{tabular}
% \end{table}
\section{Methods}

\subsection{Dataset}
In the LISA 2025 Challenge, quality assessment involves scoring the presence of seven common artifacts on pediatric brain magnetic resonance images: Banding, Contrast, Motion, Distortion, Noise, Positioning, and Zipper. Each artifact is rated on a three-point severity scale: 0 for no artifact, 1 for moderate, and 2 for severe.
The dataset provided by the challenge organizers consists of 532 brain magnetic resonance images acquired at a low magnetic field strength of 0.064T, representing 244 unique pediatric subjects. 
Each subject had up to three scans acquired in different orientations: axial, coronal, and sagittal. The severity of artifacts varied across scans.

As illustrated in Table \ref{tab:artifact_distribution}, scans containing artifacts are underrepresented. A solution proposed by the first-place winner \cite{sundaresan2024automatedqualityassessmentusing} involved increasing the proportion of scans with artifacts through simulation. Following this approach, we applied artifact simulation using TorchIO, adopting the same parameters as in previous work for all artifact types except motion. Specifically, for moderate motion (level 1), we increased the rotation severity from three to five degrees, and for severe motion (level 2), from seven to ten degrees. These adjustments resulted in more visually distinguishable motion artifacts, ensuring clearer degradation corresponding to the assigned severity level. The distribution of artifacts before and after simulation is shown in Table \ref{tab:artifact_distribution}. It is worth noting that although the number of class 1 and 2 instances increased, the overall distribution remains imbalanced. For training, scans were resized to 128×128×128, followed by augmentations such as normalization, center spatial cropping, and random rotation.

\begin{table}[t]
\centering
\caption{Distribution of artifact severity before and after simulation.}
\label{tab:artifact_distribution}
\begin{tabularx}{\textwidth}{l*{6}{>{\centering\arraybackslash}X}}
\toprule
 & \multicolumn{3}{c}{Before} & \multicolumn{3}{c}{After} \\
\cmidrule(lr){2-4} \cmidrule(lr){5-7}
Artifact & Class 0 & Class 1 & Class 2 & Class 0 & Class 1 & Class 2 \\
\midrule
Noise & 426 & 60 & 46 & 734 & 122 & 97 \\
Zipper & 398 & 105 & 29 & 686 & 201 & 66 \\
Positioning & 470 & 47 & 15 & 810 & 106 & 37 \\
Banding & 504 & 15 & 13 & 871 & 52 & 30 \\
Motion & 384 & 78 & 70 & 672 & 147 & 134 \\
Contrast & 375 & 134 & 23 & 637 & 265 & 51 \\
Distortion & 435 & 56 & 41 & 782 & 101 & 70 \\
\bottomrule
\end{tabularx}
\end{table}

\subsection{Model Description}
To predict the severity of artifacts from MRI scans, we employ a multitask learning framework. As demonstrated by \cite{kim2025axisguided}, incorporating scan plane classification as an auxiliary task enhances quality assessment, as the appearance of artifacts can vary with anatomical orientation.

In BRIQA, each input scan $\mathbf{x}$ is processed by an encoder $f_\theta(\cdot)$, which branches into two heads: one for the classification of the severity of the artifact and the other for the classification of the scan plane (axis). To mitigate the effects of class imbalance between severity levels $c' \in {0, 1, 2}$, BRIQA adopts a gradient-based loss reweighting strategy. 

For each class \( c \), BRIQA calculates how much that class contributes to the training signal by measuring the size of the gradients it produces. Specifically, BRIQA computes the \(\ell_2\) norm of the gradient of the classification loss \( \mathcal{L}_{\text{cls}}^{(c)} \) when considering only the samples that belong to class \( c \). This gradient is taken with respect to the parameters of the classification head, denoted as \( \theta_{\text{cls}} \). The result is a scalar value \( \phi_c \), which reflects the overall magnitude of the update that class \( c \) would induce on the classification head if it were trained in isolation:

\begin{equation}
\phi_c = \left\| \nabla_{\theta_{\text{cls}}} \mathcal{L}_{\text{cls}}^{(c)} \right\|_2.
\end{equation}

\noindent To rebalance the contributions from each class, BRIQA normalizes the gradients by the smallest observed norm $\ell_2$:

\begin{equation}
\alpha_c = \frac{\min_{c'} \phi_{c'}}{\phi_c}.
\end{equation}

\noindent These weights are then used to compute a weighted classification loss:

\begin{equation}
\mathcal{L}_{\text{cls}} = \sum_{c \in c'} \alpha_c \cdot \mathcal{L}_{\text{cls}}^{(c)}.
\end{equation}

\noindent Finally, the total loss for each batch is defined as:
\begin{equation}
\mathcal{L}_{\text{total}} = \mathcal{L}_{\text{cls}} + \mathcal{L}_{\text{axis}}.
\end{equation}

\noindent where $\mathcal{L}_{\text{axis}}$ encourages orientation-sensitive representations.

\noindent \textbf{Batch Configurations.} We experiment with two different configurations to form training batches to handle class imbalance and improve learning stability.

\textit{Standard Batching.} The first configuration employs standard random sampling, where training batches are formed by shuffling the dataset without enforcing any class distribution constraints.

\begin{figure}[t]
  \centering
    \includegraphics[width=\linewidth]{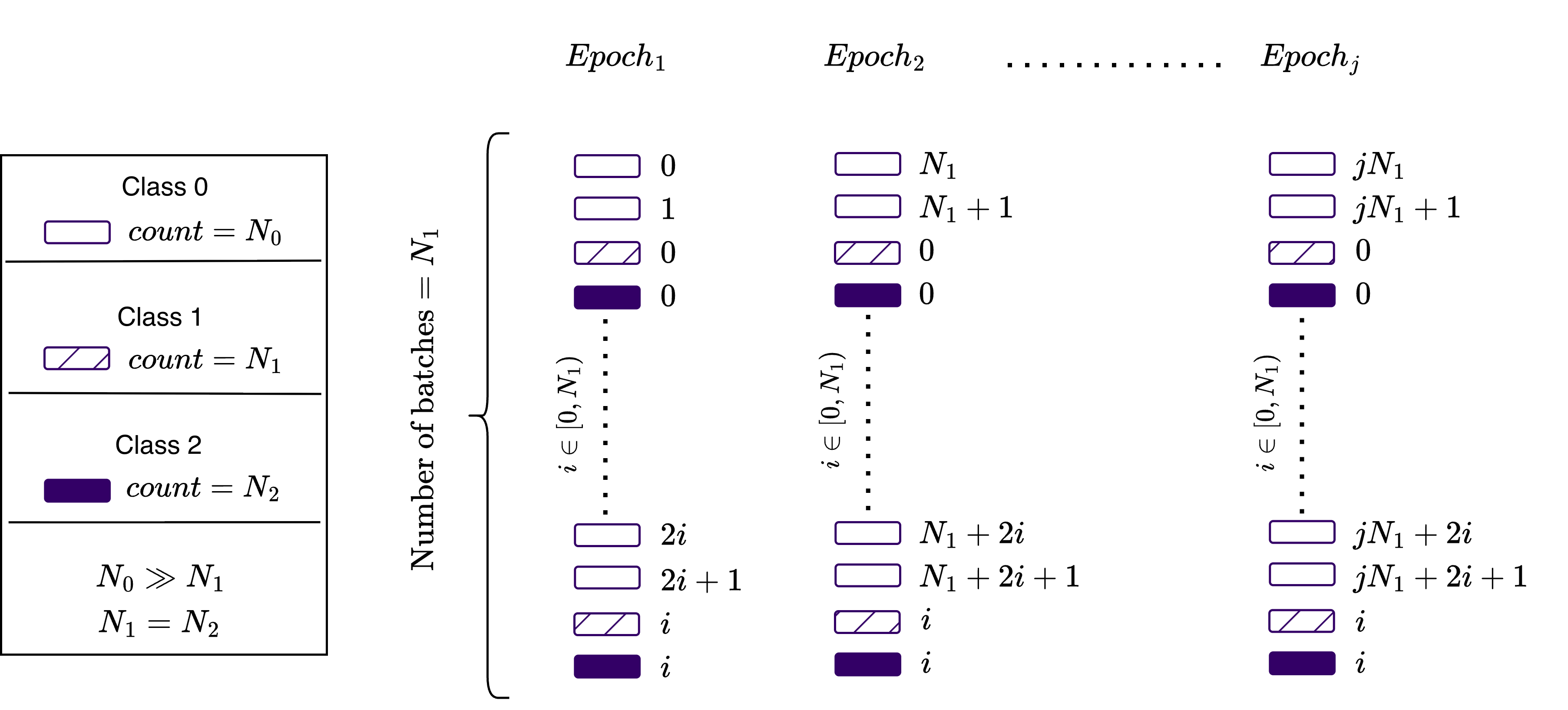}\\
    \caption{Rotating batch configuration. Samples from class~0 are drawn using a rotating buffer mechanism, where their indices are cyclically shifted across epochs according to a modular offset. Note that for each sample, we take the remainder of the calculated index with respect to $N_0$ to achieve the cycling effect; this operation is not shown in the figure.}
    \label{fig:rotating_batch}
\end{figure}

\textit{Rotating Batching.} This is a custom configuration that introduces epoch-wise variation in the selection of class 0 samples while maintaining a fixed class ratio within each batch. Each batch of size 4 contains two samples from class 0, one from class 1, and one from class 2. To address the scarcity of class 2, we apply random upsampling to match the number of class 1 samples. Unlike standard random sampling, class 0 examples are drawn using a rotating buffer strategy: their indices are cyclically shifted across epochs using a modular offset. This strategy is illustrated in Figure \ref{fig:rotating_batch}. This ensures uniform usage of all class 0 samples over time, improving training diversity. Importantly, batches are constructed before data augmentations, enabling consistent per-batch composition while introducing controlled epoch-level variation. To our knowledge, this rotating buffer mechanism is a novel contribution.
\subsection{Experimental Setup}
To evaluate BRIQA, we experimented with a range of encoder backbones, including DenseNet and ResNet variants, as well as MedNeXtS \cite{roy2024mednext}. In addition to backbone comparisons, we compare our method against \cite{zhu2025mambaout}, which considers the problem as multilabel classification. We also compare with the incorporation of a frequency-based encoder from \cite{lou2024noref}. Specifically, we applied the Discrete Fourier Transform (DFT) to the input MRI and passed the transformed image through a separate encoder. The output of the DFT-based encoder was concatenated with the features of the spatial MRI encoder before classification to capture complementary information in the frequency domain.

To analyze the impact of orientation-aware training, we compared models trained without rotation augmentation to those trained with random rotation up to 180 degrees. Additionally, we evaluated the effect of different loss functions, including Cross-Entropy (CE), Weighted Cross-Entropy, Focal Loss, and Ordinal Loss\cite{cornloss}.

The dataset was split into training and internal validation sets at the patient level, with 80\% of patients used for training and 20\% for validation. All models were trained for 150 epochs on an NVIDIA A100-SXM4-40GB GPU, using the Adam optimizer with a learning rate of \(1 \times 10^{-5}\), and a Cosine Annealing learning rate scheduler.

\section{Results}

Table \ref{tab:1} presents weighted F1 scores for detecting seven MRI artifacts in various encoders with classification heads, showing that no single architecture performs best in all types of artifacts. For example, DenseNet169 excels in detecting Zipper, Positioning, and Noise artifacts (0.844, 0.853, and 0.872, respectively), while Resnet18 achieves the highest scores on Banding, Motion and Contrast (0.947, 0.736, and 0.822, respectively). Simpler models like Resnet10 also perform competitively, particularly on Banding and Noise, outperforming deeper models in some cases. Meanwhile, MedNeXtS \cite{roy2024mednext} struggles with Motion and Distortion. The MLMambaOut \cite{zhu2025mambaout} architecture, which uses a single backbone for multi-label classification, demonstrates limited effectiveness across several artifact types regardless of the size of the model.

\begin{table}[t]
\centering
\caption{Weighted F1-scores across different MRI artifact categories using various encoder backbones and MambaOut variants. The best scores per artifact are highlighted in \textbf{bold}.}
\begin{adjustbox}{max width=\textwidth}
\begin{tabular}{lcccccccc}
\toprule
\textbf{Architecture} & \textbf{Noise} & \textbf{Zipper} & \textbf{Positioning} & \textbf{Banding} & \textbf{Motion} & \textbf{Contrast} & \textbf{Distortion} & \textbf{Mean}\\
\midrule
\multicolumn{9}{l}{\textbf{Encoder}} \\
Resnet10                  & \textbf{0.844} & 0.836 & 0.826 & \textbf{0.947} & 0.687 & 0.813 & 0.765 & \textbf{0.817} \\
Resnet18                  & 0.805 & 0.822 & 0.830 & \textbf{0.947} & \textbf{0.736} & \textbf{0.822} & 0.754 & \textbf{0.817} \\
Resnet50                  & 0.827 & 0.803 & 0.846 & \textbf{0.947} & 0.727 & 0.800 & 0.753 & 0.814 \\
Resnet101                 & 0.815 & 0.807 & 0.856 & 0.926 & 0.730 & 0.788 & 0.753  & 0.811\\
DenseNet169               & \textbf{0.844} & \textbf{0.853} & \textbf{0.872} & 0.926 & 0.716 & 0.788 & 0.776 & 0.825\\
DenseNet264               & 0.698 & \textbf{0.853} & 0.870 & 0.942 & 0.731 & 0.795 & 0.767 & 0.808 \\
MedNeXtS \cite{roy2024mednext}                & 0.807 & 0.761 & 0.789 & 0.881 & 0.486 & 0.746 & 0.601 & 0.725\\
\midrule
\multicolumn{9}{l}{\textbf{MLMambaOut}} \\
MambaOut tiny   \cite{zhu2025mambaout}         & 0.807 & 0.845 & 0.826 & 0.924 & 0.722 & 0.744 & 0.763 & 0.805\\
MambaOut small  \cite{zhu2025mambaout}         & 0.809 & 0.838 & 0.801 & 0.946 & 0.690 & 0.789 & \textbf{0.821} & 0.813\\
MambaOut base \cite{zhu2025mambaout}         & 0.811 & 0.814 & 0.834 & \textbf{0.947} & 0.698 & 0.767 & 0.790 & 0.809\\
% \midrule\midrule
% \textbf{Best combination }                    &\textbf{ 0.844}     & \textbf{0.853}     &\textbf{ 0.872}     & \textbf{0.947}     & \textbf{0.736}     & \textbf{0.822} & \textbf{0.776}     & \textbf{0.836}  \\
\bottomrule
\end{tabular}
\end{adjustbox}
\label{tab:1}
\end{table}
After selecting the best-performing backbone for each artifact, we conducted experiments under different training configurations. Table \ref{tab:2} compares the baseline setup, where reweighting is not applied, with BRIQA. In the baseline, we experiment with various loss functions, regularization techniques, and a fusion model that combines spatial MRI features with Discrete Fourier Transform (DFT) representations. 

When comparing loss functions in the baseline setting, we observe that weighted cross-entropy and ordinal loss outperform standard cross-entropy, achieving mean scores of 0.779 and 0.763, respectively, compared to 0.745 without rotation. Interestingly, applying rotation in the cross-entropy setup leads to a 0.016 improvement in the mean score. However, rotation does not consistently yield better performance. For instance, in the regularization setting, the no-rotation variant achieves a mean score of 0.780, the second highest overall, and obtains the best micro-averaged scores, outperforming the rotated version which averages 0.750. Finally, the fusion experiment, which incorporates spectral information via the DFT, shows performance close to that of the second-best configuration, suggesting that frequency-domain features may offer complementary benefits for artifact detection.

Across all configurations, BRIQA improves performance. For cross-entropy without rotation, the mean score increases from 0.745 to the highest overall score of 0.799 with rotating batching. This setup also achieves the best macro scores among all experiments while maintaining weighted and micro scores within 0.01 of the second-best results.

\begin{table}[t]
\centering

\caption{Performance metrics for different methods. \(\circ\) – No rotation (0°),  \(\circlearrowright\) – Rotation 180°. Best results are highlighted in \textbf{bold}, and second-best results are \underline{underlined}.}
\begin{adjustbox}{max width=\textwidth}
\begin{tabular}{l ccccc ccccc ccccc c}
\toprule
\textbf{Method} 
& \multicolumn{5}{c}{\textbf{Weighted}} 
& \multicolumn{5}{c}{\textbf{Macro}} 
& \multicolumn{5}{c}{\textbf{Micro}} 
& \textbf{Mean} \\
\cmidrule(lr){2-6} \cmidrule(lr){7-11} \cmidrule(lr){12-16}
& Prec. & Rec. & F1 & F2 & Acc. 
& Prec. & Rec. & F1 & F2 & Acc. 
& Prec. & Rec. & F1 & F2 & Acc. 
&  \\
\midrule
\midrule
\rowcolor{gray!15}
\multicolumn{17}{l}{\textbf{Baseline}} \\
\midrule
\multicolumn{17}{l}{\textbf{Standard Batch: Loss Variations}} \\
CE \(\circ\)          & 0.800 & 0.846 & 0.818 & 0.834 & 0.846 & 0.587 & 0.552 & 0.560 & 0.554 & 0.552 & 0.846 & 0.846 & 0.846 & 0.846 & 0.846 & 0.745 \\
CE \(\circlearrowright\)                   & 0.817 & 0.834 & 0.818  & 0.829 & 0.839 & 0.708 & 0.576 &0.619  & 0.590 & 0.576 & 0.839 & 0.840 & 0.840 & 0.840  & 0.839 & 0.761 \\
Ordinal Loss \(\circ\)                 & 0.828 & 0.841 & 0.821 & 0.831 & 0.841 & \textbf{0.741} & 0.572 & 0.607 & 0.582 & 0.572 & 0.841 & 0.841 & 0.841 & 0.841 & 0.841  & 0.763 \\
Weighted CE \(\circ\)                 & 0.834 & 0.857 & \underline{0.842} & \underline{0.851} & 0.857 & 0.661 & 0.620 & 0.629 & 0.622 & 0.620 & \underline{0.857} & \underline{0.857} &\underline{0.857} & \underline{0.857} & \underline{0.857} & 0.779 \\
\midrule
\multicolumn{17}{l}{\textbf{Standard Batch: Regularization}} \\
CE \(\circ\)              & 0.835 & \textbf{0.860 }& 0.842 & \textbf{0.852} & \textbf{0.860} & 0.660 & 0.621 & 0.628 & 0.623 & 0.621 & \textbf{0.860} & \textbf{0.860} & \textbf{0.860} & \textbf{0.860 }& \textbf{0.860} & \underline{0.780} \\
CE \(\circlearrowright\)           & 0.825 & 0.815 & 0.810 & 0.812 & 0.815 & 0.670 & 0.607 & 0.608 & 0.605 & 0.607 & 0.815 & 0.815 & 0.815 & 0.815 & 0.815 & 0.750 \\
\midrule
\multicolumn{17}{l}{\textbf{Standard Batch: DFT Fusion}} \\
CE \(\circ\)           & 0.829 & 0.845 & 0.834 & 0.840 & 0.845 & 0.701 & 0.631 & 0.659 & 0.641 & 0.631 & 0.845 & 0.845 & 0.845 & 0.845 & 0.845 & 0.779 \\
\midrule
\midrule
\rowcolor{gray!15}
\multicolumn{17}{l}{\textbf{BRIQA}} \\
\midrule
\multicolumn{17}{l}{\textbf{Standard Batch: Loss Variations}}\\
CE \(\circ\)       & \underline{0.840} & \underline{0.853} & 0.844 & 0.849 & 0.853 & \underline{0.732} & 0.657 & 0.688 & 0.668 & 0.657 & 0.853 & 0.853 & 0.853 & 0.853 & 0.853 & 0.794 \\
CE \(\circlearrowright\)     & 0.789 & 0.743 & 0.763 & 0.750 & 0.743 & 0.576 & 0.612 & 0.592 & 0.607 & 0.619 & 0.743 & 0.743 & 0.743 & 0.743 & 0.743 & 0.701 \\
Ordinal Loss \(\circ\)& 0.838 & 0.786 & 0.801 & 0.789 & 0.786 & 0.653 & 0.649 & 0.621 & 0.625 & 0.649 & 0.786 & 0.786 & 0.786 & 0.786 & 0.786 & 0.742 \\
Focal loss \(\circ\) & 0.818 & 0.818 & 0.818 & 0.818 & 0.818 & 0.642 & 0.648 & 0.645 & 0.647 & 0.648 & 0.818 & 0.818 & 0.818 & 0.818 & 0.818 & 0.760 \\
\midrule
\multicolumn{17}{l}{\textbf{Standard Batch: DFT Fusion}} \\
CE \(\circ\) & 0.821 & 0.827 & 0.823 & 0.825 & 0.827 & 0.665 & 0.683 & 0.671 & \underline{0.677} & 0.683 & 0.827 & 0.827 & 0.827 & 0.827 & 0.827 & 0.776 \\
% \midrule
% \multicolumn{17}{l}{\textbf{Ours: Fixed-Ratio Batching}} \\
% Baseline \(\circ\)       & 0.816 & 0.777 & 0.790 & 0.779 & 0.777 & 0.584 &\underline{ 0.684} & 0.6070 & 0.642 & \underline{0.684} & 0.777 & 0.777 & 0.777 & 0.777 & 0.777 & 0.735 \\
% Baseline \(\circlearrowright\)     & 0.821 & 0.812 & 0.816 & 0.814 & 0.812 & 0.665 & 0.678 & \underline{0.671} & \underline{0.677} & 0.677 & 0.812 & 0.812 & 0.812 & 0.812 & 0.812 & 0.767 \\
\midrule
\multicolumn{17}{l}{\textbf{Rotating Batch: BRIQA}} \\
CE \(\circ\)       & \textbf{0.843} & 0.849 & \textbf{0.846} & 0.848 & 0.849 & 0.724 & \textbf{0.690 }& \textbf{0.706} & \textbf{0.696} & \textbf{0.690} & 0.849 & 0.849 & 0.849 & 0.849 & 0.849  &\textbf{ 0.799} \\
\bottomrule
\end{tabular}
\end{adjustbox}
\label{tab:2}

\end{table}

To better understand where these gains are most impactful, Table~\ref{tab:3} presents a detailed breakdown of performance improvements in the seven types of MRI artifacts. The model demonstrates consistent improvements in most artifacts, particularly in Noise, Zipper, and Distortion, where all metrics show notable gains. For example, noise and distortion exhibit substantial increases in macro F1 and F2 scores, indicating enhanced sensitivity to rare or harder-to-classify severity levels. Although Banding achieved the highest weighted and micro F1 scores (0.919 and 0.905, respectively), its macro performance was slightly lower than that of other artifacts, probably due to its prevalence in the dataset. Zipper achieved the highest macro F1 score, improving by nearly 10\% over baseline.

\begin{table}[t]
\centering
\caption{Performance metrics of the best-performing model across artifact types. \((\uparrow)\) indicates improvement over the CE without gradient reweighting, and \((\downarrow)\) denotes a performance decrease.}
\begin{adjustbox}{max width=\textwidth}
\begin{tabular}{l@{\hskip 10pt}c@{\hskip 10pt}c@{\hskip 10pt}c@{\hskip 10pt}c@{\hskip 10pt}c@{\hskip 10pt}c@{\hskip 10pt}c}

\toprule
\textbf{Metric} 
& \textbf{Noise} & \textbf{Zipper} & \textbf{Positioning} 
& \textbf{Banding} & \textbf{Motion} & \textbf{Contrast} & \textbf{Distortion} \\
\midrule

\multicolumn{8}{l}{\textbf{Weighted}} \\
Precision      & $0.863_{\uparrow0.238}$ & $0.871_{\uparrow0.021}$ & $0.902_{\uparrow0.037}$ & 
$0.936_{\uparrow0.001}$ & 
$0.755_{\uparrow0.025}$& 
$0.811_{\downarrow0.012}$& 
$0.870_{\uparrow0.100}$\\
Recall         & $0.876_{\uparrow0.086}$ & $0.867_{\uparrow0.010}$ & 
$0.838_{\downarrow0.048}$ & 
$0.905_{\downarrow0.057}$ & 
$0.781_{\uparrow0.029}$& 
$0.810_{\downarrow0.019}$& 
$0.867_{\uparrow0.057}$\\
F1-score       & $0.865_{\uparrow0.167}$ & $0.863_{\uparrow0.010}$ & 
$0.860_{\downarrow0.011}$ & 
$0.919_{\downarrow0.028}$ & 
$0.747_{\uparrow0.012}$ & 
$0.799_{\downarrow0.023}$& 
$0.861_{\uparrow0.085}$\\
F2-score       & $0.871_{\uparrow0.120}$ & $0.864_{\uparrow0.009}$ & 
$0.844_{\downarrow0.036}$ & 
$0.910_{\downarrow0.045}$ & 
$0.765_{\uparrow0.020}$& 
$0.803_{\downarrow0.022}$&
$0.863_{\uparrow0.069}$\\
Accuracy       & $0.876_{\uparrow0.086}$ & $0.867_{\uparrow0.010}$ & 
$0.838_{\downarrow0.048}$ &
$0.905_{\downarrow0.057}$ & 
$0.781_{\uparrow0.029}$& 
$0.810_{\downarrow0.019}$& 
$0.867_{\uparrow0.057}$\\
\hline
\multicolumn{8}{l}{\textbf{Macro}} \\
Precision      & $0.747_{\uparrow0.483}$ & $0.865_{\uparrow0.214}$ &
$0.713_{\uparrow0.037}$ & 
$0.593_{\downarrow0.061}$ & 
$0.716_{\uparrow0.080}$& 
$0.705_{\downarrow0.036}$& 
$0.725_{\uparrow0.177}$\\
Recall         & $0.728_{\uparrow0.396}$ & $0.684_{\uparrow0.055}$ & 
$0.794_{\uparrow0.177}$  & 
$0.643_{\uparrow0.088}$ & 
$0.598_{\uparrow0.013}$& 
$0.738_{\uparrow0.030}$& 
$0.636_{\uparrow0.215}$\\
F1-score       & $0.725_{\uparrow0.430}$ & $0.731_{\uparrow0.098}$ & 
$0.732_{\uparrow0.097}$ & 
$0.605_{\uparrow0.012}$ & 
$0.625_{\uparrow0.022}$& 
$0.698_{\downarrow0.022}$& 
$0.657_{\uparrow0.217}$\\
F2-score       & $0.725_{\uparrow0.408}$ & $0.698_{\uparrow0.068}$ & 
$0.761_{\uparrow0.138}$& 
$0.622_{\uparrow0.053}$ & 
$0.605_{\uparrow0.014}$& 
$0.716_{\uparrow0.004}$& 
$0.641_{\uparrow0.216}$\\
Accuracy       & $0.728_{\uparrow0.395}$ & $0.684_{\uparrow0.055}$ & 
$0.793_{\uparrow0.177}$ &
$0.643_{\uparrow0.088}$ & 
$0.598_{\uparrow0.013}$& 
$0.738_{\uparrow0.030}$& 
$0.636_{\uparrow0.215}$\\
\hline
\multicolumn{8}{l}{\textbf{Micro}} \\
Precision      & $0.876_{\uparrow0.086}$ & $0.867_{\uparrow0.010}$ & 
$ 0.838_{\downarrow0.048}$ & 
$0.905_{\downarrow0.057}$ & 
$0.781_{\uparrow0.029}$& 
$0.810_{\downarrow0.019}$& 
$0.867_{\uparrow0.057}$\\
Recall         & $0.876_{\uparrow0.086}$ & $0.867_{\uparrow0.010}$ & 
$0.838_{\downarrow0.048}$  & 
$0.905_{\downarrow0.057}$ & 
$0.781_{\uparrow0.029}$& 
$0.810_{\downarrow0.019}$& 
$0.867_{\uparrow0.057}$\\
F1-score       & $0.876_{\uparrow0.086}$ & $0.867_{\uparrow0.010}$ & 
$0.838_{\downarrow0.048}$ & 
$0.905_{\downarrow0.057}$ & 
$0.781_{\uparrow0.029}$& 
$0.810_{\downarrow0.019}$& 
$0.867_{\uparrow0.057}$\\
F2-score       & $0.876_{\uparrow0.086}$ & $0.867_{\uparrow0.010}$ & 
$0.838_{\downarrow0.048}$ & 
$0.905_{\downarrow0.057}$ & 
$0.781_{\uparrow0.029}$& 
$0.810_{\downarrow0.019}$& 
$0.867_{\uparrow0.057}$\\
Accuracy       & $0.876_{\uparrow0.086}$ & $0.867_{\uparrow0.010}$ & 
$0.838_{\downarrow0.048}$ & $0.905_{\downarrow0.057}$& 
$0.781_{\uparrow0.029}$& 
$0.810_{\downarrow0.019}$& 
$0.867_{\uparrow0.057}$\\
\hline
{\textbf{Mean}} 
& $0.826_{\uparrow0.216}$ & $0.822_{\uparrow0.040}$ & $0.818_{\uparrow0.019}$ & 
$0.814_{\downarrow0.020}$ & 
$0.725_{\uparrow0.027}$& 
$0.778_{\downarrow0.012}$& 
$0.797_{\uparrow0.113}$ \\
\bottomrule
\end{tabular}
\end{adjustbox}
\label{tab:3}
\end{table}

\section{Discussion}
\textbf{Is One Backbone Architecture Enough?} Performance in Table~\ref{tab:1} suggests that a one-size-fits-all architecture may not be ideal for MRI artifact detection. Instead, leveraging the complementary strengths of diverse backbones could offer improved robustness across artifact types. The consistent variability in per-artifact performance across architectures, especially for more challenging categories like Distortion and Motion, indicates that certain models are more sensitive to specific artifact patterns. Rather than seeking a universally strong backbone, it may be more effective to utilize this diversity and design adaptive frameworks that combine multiple models to capitalize on their respective strengths. We hypothesize that the performance difference comes from how each network propagates features. ResNet adds features through residual connections, capturing global structure and performing better on artifacts affecting the whole image, like motion, contrast, and banding. DenseNet concatenates features, preserving fine detail, which helps with localized or textural artifacts such as zipper lines and positioning shifts.

\noindent \textbf{Is Cross-Entropy Enough?}
The baseline experiments show that standard cross-entropy loss is suboptimal when compared to both weighted cross-entropy and ordinal loss. These alternative loss functions are more effective in addressing label imbalance and capturing ordinal relationships between classes, resulting in higher macro- and weighted scores. However, when paired with BRIQA, the standard cross-entropy loss achieves the highest overall performance compared to other losses.  This improvement comes from the rotating batch configuration’s ability to expose the model to diverse artefact combinations across training iterations,  helping the model generalise better across underrepresented classes. In addition to weighting the loss based on gradient contributions, which eliminates the need for explicit reweighting mechanisms in focal loss.
In this context, standard cross-entropy benefits from a more uniform and representative training distribution, making it competitive, even surpassing more specialized loss functions. 
 
\noindent \textbf{Does Scan Rotation Always Help?}
If rotation-based augmentation is applied, the benefits appear inconsistent in different settings. While it yields marginal gains in some configurations at baseline, such as standard cross-entropy, it can degrade performance in more optimized settings like regularized training, where the non-rotated variant achieved a substantially better mean (0.780 vs. 0.750) and the highest micro-average overall. In contrast to baseline, in batch configuration settings, applying rotation with standard cross entropy degraded the performance. This suggests that rotation may inject noise or disrupt spatial integrity in some representations, especially when models already have strong regularisation or batching restrictions.
 
\noindent \textbf{Can Custom Batching Improve Learning from Imbalanced Data?} Our findings highlight the significant impact of batching design on model performance under class imbalance. Unlike standard batching, which randomly samples data and may repeatedly draw from overrepresented classes, rotating batching enforces a fixed class ratio within each batch while systematically cycling through majority class (class 0) samples across epochs. This strategy ensures that minority classes are consistently represented in every batch, while the majority class is varied to maintain diversity and prevent oversaturation. By balancing the gradient signal across classes, rotating batching helps stabilize training and reduces the tendency to overfit to dominant class patterns. This is particularly important in the context of artifact detection, where severe artifact cases (class 2) are underrepresented. Without careful batching, the model could learn to ignore rare artifacts in favor of more frequent, clean scans. Rotating batching ensures that learning remains attentive to all severity levels, improving generalization and classification robustness.

\noindent \textbf{Does Frequency Domain Help?}
The observed performance of the DFT fusion setup suggests that incorporating frequency-domain features offers complementary benefits to spatial representations. Although the mean score of the DFT fusion model falls slightly below that of the best regularized configuration, its strong performance in macro-averaged metrics indicates improved generalization to underrepresented severity levels. This is particularly relevant for artifacts, where frequency patterns may be more informative than spatial textures alone. However, the relatively modest gains compared to those of other setups suggest that a simple fusion strategy may not fully exploit the potential of spectral information. More advanced integration mechanisms, such as attention-based fusion, may be necessary to effectively combine spatial and frequency-domain features.

\noindent \textbf{Computational Requirements.} 
BRIQA takes 0.008–0.016 s per sample with 740 MB peak GPU memory for DenseNet-based variants (Noise, Zipper, Positioning), 0.021 s and 4.3 GB for ResNet18 models (Banding, Contrast, Motion), and 0.041 s with 7.2 GB for the MedNextS Distortion model.

\section{Conclusion}
In this work, we addressed the challenge of automatic classification of MRI artifact severity under class imbalance by proposing BRIQA, which integrates axis prediction, gradient-based loss reweighting, and a rotation-based batch construction strategy. Our findings show that architectural diversity can be leveraged for better performance across different artifact categories, while rotating batching significantly enhances generalization by ensuring consistent exposure to minority classes. Future work may explore dynamic ensemble methods based on artifact type or severity distribution, as well as artifact-specific expert models trained to handle visually distinct patterns. While BRIQA demonstrates improved performance, several limitations warrant consideration. First, the multi-architecture approach requires training and maintaining multiple models, increasing computational overhead compared to single-model solutions. Second, the relatively small dataset size and reliance on simulated artifacts may limit generalization to other low-field MRI systems or diverse patient populations. Clinical validation on larger, multi-center datasets is necessary to confirm BRIQA's utility in real diagnostic workflows.

\subsubsection{\discintname}
The authors have no competing interests.
\bibliographystyle{plain}  %
\bibliography{references} 

\begin{thebibliography}{10}

\bibitem{introduction_lmri_dis_and_advantages}
Thomas~Campbell Arnold, Colbey~W Freeman, Brian Litt, and Joel~M Stein.
\newblock Low-field mri: clinical promise and challenges.
\newblock {\em Journal of Magnetic Resonance Imaging}, 57(1):25--44, 2023.

\bibitem{kim2025axisguided}
Hyunwook Kim, Jinew Seo, Seiyoung Ryu, Joon~Hyung Park, Sungchul On, and Jinwha Choi.
\newblock Axis-guided quality assessment and multi-label hippocampal and ventricular segmentation in low-resolution pediatric brain mri.
\newblock In Natasha Lepore and Marius~George Linguraru, editors, {\em Proceedings of the Low Field Pediatric Brain Magnetic Resonance Image Segmentation and Quality Assurance (LISA 2024)}, volume 15515 of {\em Lecture Notes in Computer Science}, pages 53--62. Springer, Cham, 2025.

\bibitem{introduction_Mri_in_neurodevelopment_2_and_non_ionizing}
Rhoshel~K Lenroot and Jay~N Giedd.
\newblock Brain development in children and adolescents: insights from anatomical magnetic resonance imaging.
\newblock {\em Neuroscience \& biobehavioral reviews}, 30(6):718--729, 2006.

\bibitem{challenge_file}
Natasha Lepore and Marius~George Linguraru.
\newblock Low field pediatric brain magnetic resonance image segmentation and quality assurance: First miccai challenge, lisa 2024, held in conjunction with miccai 2024, marrakesh, morocco, october 10, 2024, proceedings, 2025.

\bibitem{lou2024noref}
Yiwei Lou, Jiayu Zhang, Dexuan Xu, Yongzhi Cao, Hanpin Wang, and Yu~Huang.
\newblock No-reference mri quality assessment via contrastive representation: Spatial and frequency domain perspectives.
\newblock In {\em 2024 IEEE International Conference on Multimedia and Expo (ICME)}, pages 1--6. IEEE, 2024.

\bibitem{resnet_better_than_densenet}
I~Putu Gede Yoga~Pramana Putra, Ni~Wayan Jeri~Kusuma Dewi, Putu Surya~Wedra Lesmana, I~Gede~Totok Suryawan, and Putu Satria~Udyana Putra.
\newblock Comparison of resnet-50 and densenet-121 architectures in classifying diabetic retinopathy.
\newblock {\em Indonesian Journal of Data and Science}, 6(1):64--72, 2025.

\bibitem{roy2024mednext}
Saikat Roy, Gregor Koehler, Constantin Ulrich, Michael Baumgartner, Jens Petersen, Fabian Isensee, Paul~F. J{\"a}ger, and Klaus~H. Maier-Hein.
\newblock Mednext: Transformer-driven scaling of convnets for medical image segmentation.
\newblock In Hayit Greenspan, Anant Madabhushi, Parvin Mousavi, Septimiu Salcudean, James Duncan, Tanveer Syeda-Mahmood, and Russell Taylor, editors, {\em Medical Image Computing and Computer Assisted Intervention -- MICCAI 2023}, pages 405--415, Cham, 2023. Springer Nature Switzerland.

\bibitem{introduction_Mri_in_diagnosis}
Uroosa Saman, Anwarul Haque, Namaya Hussain, and Bushra Shamim.
\newblock Utility of magnetic resonance imaging of brain in neurocritically ill children in pediatric intensive care unit: A single-center retrospective observational study.
\newblock {\em Journal of Pediatric Critical Care}, 11(1):6--9, 2024.

\bibitem{FetMRQC}
Thomas Sanchez, Oscar Esteban, Yvan Gomez, Elisenda Eixarch, and Meritxell~Bach Cuadra.
\newblock Fetmrqc: automated quality control for fetal brain mri.
\newblock pages 3--16, 2023.

\bibitem{cornloss}
Xintong Shi, Wenzhi Cao, and Sebastian Raschka.
\newblock Deep neural networks for rank-consistent ordinal regression based on conditional probabilities.
\newblock {\em Pattern Analysis and Applications}, 26(3):941--955, 2023.

\bibitem{sundaresan2024automatedqualityassessmentusing}
Vaanathi Sundaresan and Nicola~K Dinsdale.
\newblock Automated quality assessment using appearance-based simulations and hippocampus segmentation on low-field paediatric brain mr images.
\newblock pages 41--52, 2024.

\bibitem{Densenet_better_than_resnet}
Tomoki Uemura, Janne~J N{\"a}ppi, Toru Hironaka, Hyoungseop Kim, and Hiroyuki Yoshida.
\newblock Comparative performance of 3d-densenet, 3d-resnet, and 3d-vgg models in polyp detection for ct colonography.
\newblock In {\em Medical Imaging 2020: computer-aided diagnosis}, volume 11314, pages 736--741. SPIE, 2020.

\bibitem{introduction_Mri_in_neurodevelopment}
Firehiwot Workneh, Theresa~Inez Chin, Kalkidan Yibeltal, Krysten North, Nebiyou Fasil, Workagegnhu Tarekegn, Betelhem~Haimanot Abate, Sarem Mulugeta, Gellila Asmamaw, Atsede Teklehaimanot, et~al.
\newblock Feasibility and acceptability of magnetic resonance imaging and electroencephalography for child neurodevelopmental research in rural ethiopia.
\newblock {\em Frontiers in Public Health}, 13:1551982, 2025.

\bibitem{bigger_not_better}
Yuan Yang, Lin Zhang, Mingyu Du, Jingyu Bo, Haolei Liu, Lei Ren, Xiaohe Li, and M~Jamal Deen.
\newblock A comparative analysis of eleven neural networks architectures for small datasets of lung images of covid-19 patients toward improved clinical decisions.
\newblock {\em Computers in Biology and Medicine}, 139:104887, 2021.

\bibitem{Zhang2024}
Wenhao Zhang, Xin Zhang, Lingyi Li, Lufan Liao, Fenqiang Zhao, Tao Zhong, Yuchen Pei, Xiangmin Xu, Chaoxiang Yang, He~Zhang, et~al.
\newblock A joint brain extraction and image quality assessment framework for fetal brain mri slices.
\newblock {\em NeuroImage}, 290:120560, 2024.

\bibitem{zhu2025mambaout}
Yueyue Zhu, Haotian Jiang, Rongqing Cai, and Geng Chen.
\newblock Multi-label mambaout for quality assessment of low-field pediatric brain mr images.
\newblock In {\em MICCAI Challenge on Low Field Pediatric Brain Magnetic Resonance Image Segmentation and Quality Assurance}, pages 3--11. Springer Nature Switzerland Cham, 2024.

\end{thebibliography}
\end{document}